\documentclass[aps,preprint,superscriptaddress]{revtex4}
\usepackage{graphicx}
\usepackage{color}
\usepackage{ulem}

\begin{document}

\title{Resonance from antiferromagnetic spin fluctuations for superconductivity in UTe$_2$}

\author{Chunruo Duan}
\affiliation{Department of Physics and Astronomy, Rice Center for Quantum Materials, Rice University, Houston, TX 77005, USA}
\author{R. E. Baumbach}
\affiliation{National High Magnetic Field Laboratory, Florida State University, Tallahassee, FL 32306, USA}
\affiliation{Department of Physics, Florida State University, Tallahassee, FL 32306, USA}
\author{Andrey Podlesnyak}
\affiliation{Neutron Scattering Sciences Division, Oak Ridge National Laboratory, Oak Ridge, TN 37831, USA}
\author{Yuhang Deng}
\author{Camilla Moir}
\author{Alexander J. Breindel}
\author{M. Brian Maple}
\affiliation{Department of Physics, University of California, San Diego, CA 92093, USA}
\author{E. M. Nica}
\affiliation{Department of Physics, Arizona State University, Tempe, AZ 85287-1504, USA}
\author{Qimiao Si}
\affiliation{Department of Physics and Astronomy, Rice Center for Quantum Materials, Rice University, Houston, TX 77005, USA}
\author{Pengcheng Dai}
\email{pdai@rice.edu}
\affiliation{Department of Physics and Astronomy, Rice Center for Quantum Materials, Rice University, Houston, TX 77005, USA}
\maketitle 

{\bf Superconductivity originates from the formation of bound (Cooper) pairs of electrons that can move through the lattice without resistance below the superconducting transition temperature $T_c$ \cite{bcs}. Electron Cooper pairs in most superconductors form anti-parallel spin singlets with total spin $S=0$ \cite{scalapinormp}, although they can also form parallel spin-triplet Cooper pairs with $S=1$ and an odd parity wavefunction \cite{andy}.  Spin-triplet pairing is important because it can host topological states and Majorana fermions relevant for quantum computation \cite{MSato2017,Kitaev2001}. Because spin-triplet pairing is usually mediated by ferromagnetic (FM) spin fluctuations \cite{andy}, uranium based materials near an FM instability are considered to be ideal candidates for realizing spin-triplet superconductivity \cite{Aoki2019a}. Indeed, UTe$_2$, which has a $T_c\approx 1.6$ K \cite{Ran2019,Aoki2019}, has been identified as a candidate for a chiral spin-triplet topological superconductor near an FM instability \cite{Ran2019,Aoki2019,Ran2019b,Knebel2019,Sundar2019,Jiao2020,Nakamine2021,Hayes2020}, although it also has antiferromagnetic (AF) spin fluctuations \cite{Thomas2020,Duan2020}.  Here we use inelastic neutron scattering (INS) to show that superconductivity in UTe$_2$ is coupled to a sharp magnetic excitation, termed resonance \cite{mignod,Wilson2006,Dai,Sato2001,Bernhoeft2000,CStock2008,OStockert2011}, at the Brillouin zone boundary near AF order.  Because the resonance has only been found in spin-singlet unconventional superconductors near an AF instability \cite{mignod,Wilson2006,Dai,Sato2001,Bernhoeft2000,CStock2008,OStockert2011}, its observation in UTe$_2$ suggests that AF spin fluctuations may also induce spin-triplet pairing \cite{Kuwabara2000} or that electron pairing in UTe$_2$ has a spin-singlet component. 
}

In conventional Bardeen-Cooper-Schrieffer superconductors, electron-lattice coupling binds electrons into spin-singlet pairs below $T_c$ without involving magnetism \cite{bcs}. In most unconventional superconductors, the proximity of superconductivity to static AF ordered states suggests AF spin fluctuations as a common thread that can pair electrons into spin singlets for superconductivity \cite{scalapinormp}. For spin-triplet candidate heavy-fermion superconductors such as UGe$_2$, URhGe, and UCoGe, superconductivity arises through suppression of the static FM order or coexists with static FM order \cite{Aoki2019a}. In unconventional spin-singlet superconductors, the resonance is a sharp magnetic excitation near an AF ordering wavevector in the superconducting state that peaks at a well defined energy $E_r$ and an intensity that tracks the superconducting order parameter \cite{Eschrig,GYu2009}. Within the weak-coupling theory of superconductivity, the resonance is a bound state inside the particle-hole continuum gap, referred to as a spin exciton, that arises from quasiparticle excitations that connect parts of the Fermi surfaces exhibiting a sign change in the superconducting order parameter  [$\Delta({\bf k})=-\Delta({\bf k}+{\bf Q})$, where $\Delta({\bf k})$ is the momentum(${\bf k}$)-dependent superconducting gap and ${\bf Q}$ is the momentum transfer connecting the two gapped Fermi surfaces] \cite{scalapinormp,Eschrig}.  In this picture, the energy of the resonance is below the sum of the energies of the superconducting gaps of the two connecting Fermi surfaces, and its wavevector dependence contains signatures of the superconducting gap symmetry \cite{scalapinormp,Eschrig}.

For uranium based heavy-fermion superconductors near a FM instability \cite{Aoki2019a}, although previous INS experiments have found FM spin fluctuations, there is no evidence that these fluctuations are coupled to superconductivity \cite{Huxley2003,Stock2011}. Similarly, although incommensurate and FM spin fluctuations were found in the spin-triplet candidate superconductor Sr$_2$RuO$_4$, they do not couple to superconductivity, and therefore suggest that spin fluctuations alone are not sufficient to induce spin-triplet superconductivity \cite{Kunkem2017,Steffens2019}. These results are consistent with nuclear magnetic resonance (NMR) Knight shift measurements that indicate that superconductivity in Sr$_2$RuO$_4$ cannot arise from a pure spin-triplet pairing state \cite{Pustogow2019}. Finally, for the spin-triplet superconductor candidate UPt$_3$ \cite{Joynt2002}, superconductivity appears to couple to very weak static AF order instead of FM spin fluctuations \cite{Aeppli1988}. Therefore, there is no experimental evidence that  superconductivity is coupled with FM spin fluctuations in any of these spin-triplet candidate materials \cite{Huxley2003,Stock2011,Kunkem2017,Steffens2019,Pustogow2019,Joynt2002,Aeppli1988}. 

We chose to study spin excitations in UTe$_2$ using INS because this technique can probe both FM and AF spin fluctuations and the effect of superconductivity on these excitations [Figs. 1(a,b)] \cite{Dai}. UTe$_2$ sits at the paramagnetic end of a series of FM heavy-fermion superconductors \cite{Ran2019,Aoki2019}, and is believed to be a spin-triplet superconductor for the following reasons: (1) Upper critical fields $H_{C2}$ that exceed the Pauli limits along all crystallographic directions \cite{Ran2019b,Knebel2019}; (2) Muon spin relaxation/rotation measurements of coexisting FM spin fluctuations and superconductivity \cite{Sundar2019}; (3) Scanning tunnelling microscopy evidence of chiral-triplet topological superconductivity \cite{Jiao2020}; (4) Exclusion of spin-singlet pairing from the $^{125}$Te Knight shifts reduction below $T_c$ measured by NMR \cite{Nakamine2021}; and (5) Breaking of time reversal symmetry below $T_c$ from a non-zero polar Kerr effect and evidence for two superconducting transitions in the specific heat \cite{Hayes2020}. Although these reasons provide circumstantial evidence for spin-triplet superconductivity, they are not conclusive proof that superconductivity in UTe$_2$ must be in a pure spin-triplet $p$-wave state. For example, although time-reversal symmetry breaking is seen by a non-zero Kerr effect, it is not confirmed by muon spin relaxation/rotation measurements; however, reasons why this might not have been visible have been discussed \cite{Sundar2019,Hayes2020}. Moreover, interpretation of the Knight shift data from NMR measurements can be ambiguous because the NMR signal only probes within a London penetration depth of the surface and therefore may not reflect bulk behavior \cite{gannon2017}. 

On the other hand, there are indications that UTe$_2$ is near an AF instability instead of an FM order \cite{Thomas2020,Duan2020}. In particular, our previous INS experiments within the $[0,K,L]$ scattering plane of UTe$_2$ reveal spin fluctuations at the incommensurate wavevectors ${\bf Q}=(0,\pm(K+ 0.57),0)$ ($K=0,1$) not far away from the Brillouin zone boundary \cite{Duan2020}. The magnetic scattering is centered around $L=0$ and dispersionless along the $L$-direction, suggesting that spin fluctuations in UTe$_2$ are two-dimensional in the $[H,K,0]$ plane [Figs. 1(c,d)]. Nevertheless, there is no evidence that they are coupled to superconductivity \cite{Duan2020}.

\noindent
{\bf Experimental Data} \newline
Here we use INS to map out the spin excitations in UTe$_2$ in the $[H,K,0]$ plane and show that superconductivity induces a resonance near the AF wavevector at an energy $E_r=7.9k_BT_c$ ($k_B$, Boltzmann's constant) and opens a spin gap at energies below the mode, analogous to what occurs in unconventional spin-singlet superconductors \cite{mignod,Wilson2006,Dai,Sato2001,Bernhoeft2000,CStock2008,OStockert2011}. Figure 1(a) shows the orthorhombic unit cell of UTe$_2$ (space group $Immm$) \cite{Ran2019}. The bulk superconductivity of our samples is confirmed by heat capacity measurements showing $T_c\approx 1.6$ K [Fig. 1(b)]. Figure 1(c) shows Brillouin zones in reciprocal space within the $[H,K,0]$ plane, where solid red ellipses (Y1,Y2,Y3) and green dots (T1,T2) are positions of spin excitations as a function of increasing energy [Fig. 1(d)]. The blue solid dots are $\Gamma$ points and nuclear Bragg peaks are at $(\pm1,\pm1)$. The energy dependence of the imaginary part of the local dynamical susceptibility $\chi^{\prime\prime}(E)$ near Y1, defined as $\chi^{\prime\prime}(E)=\int_{\rm BZ}\chi^{\prime\prime}({\bf Q},E)d{\bf Q}/\int_{\rm BZ}d{\bf Q}$ within a Brillouin zone where $E$ is the excitation energy \cite{Dai}, above and below $T_c$ reveals a clear resonance and a spin gap in the superconducting state [Fig. 1(e)].  On the other hand, $\chi^{\prime\prime}(E)$ near T1 shows no observable changes across $T_c$ [Fig. 1(f)]. Figures 1(g) and 1(h) compare the energy of the resonance mode with unconventional spin-singlet superconductors \cite{GYu2009}, indicating that the mode deviates from the current trend for these materials.

Figure 2(a-f) shows the wavevector dependence of elastic and inelastic scattering in UTe$_2$ as a function of increasing energy at base temperature (BT$=0.25$ K) and above $T_c$ ($T=2$ K). In the elastic channel, we find nuclear Bragg peaks at the $(0,-2,0)$ and $(1,\pm1,0)$ positions and no evidence of magnetic order at BT [Fig. 2(a), Extended Data Fig. 4(a,b)]. On increasing energy to $E=0.4\pm 0.1$ meV, there is clear scattering at the Brillouin zone boundary position (Y1 point) in the normal state that is suppressed at BT [Fig. 2(b)]. Upon further increasing energies to $E=0.7\pm0.1$, $1.0\pm 0.1$, $1.5\pm 0.1$, $2.0\pm 0.1$ meV, spin excitations are still well defined along the $[0,K,0]$ direction at Y points but broaden progressively along the $[H,0,0]$ direction [Figs. 2(c-f)]. In addition, we see clear magnetic scattering at T points of reciprocal space for energies above $E=0.7\pm0.1$ meV [Figs. 2(c-f)]. Although the spin excitation intensity increases below $T_c$ at $E=1.0\pm 0.1$ meV for all equivalent Y points  [Fig. 1(d)], they are virtually temperature independent across $T_c$ at Y points for energies above 1.3 meV and at T points for all energies.

Figures 3(a) and 3(b) summarize the evolution of spin excitations for energies above $E=2.1$ meV at BT.  At $E=3.25\pm 0.25$ meV, spin excitations are still well-defined along the $[0,K,0]$ direction but extend to the entire Brillouin zone boundary along the $[H,0,0]$ direction [Fig. 3(a)]. Finally, at $E=5.25\pm 0.25$ meV, they become weak and diffusive, but still center around the Brillouin zone boundary broadly along the $[H,0,0]$ direction [Fig. 3(b)]. The ${\bf Q}$-$E$ map along the $[0,K,0]$ direction reveals clear spin excitations stemming from Y points that disappear above 7 meV [Fig. 3(c)].  The temperature dependence of the scattering along the $[0,K,0]$ direction across $T_c$ is shown in Figs. 3(d) and 3(e), where the superconductivity-induced spin gap and resonance are observed at the Y1 and Y2 points. The broad dispersive scattering from the $(0,-2,0)$ nuclear Bragg peak is due to a temperature-independent acoustic phonon \cite{Duan2020}. Figure 3(f) shows the ${\bf Q}$-$E$ map along the $[0.5,K,0]$ direction. We see clear rod-like magnetic scattering stemming from the T points in reciprocal space above $E=0.5$ meV [Fig. 3(f)], but these excitations do not respond to superconductivity (Extended Data Fig.7).

To further demonstrate that spin excitations at the Y1 position are coupled to superconductivity, we carried out high resolution measurements using an incident neutron energy of $E_i=2.5$ meV. The wavevector dependent scattering at $E=0.275\pm 0.025$ meV below $T_c$ [Fig. 4(a)] and above [Fig. 4(b)] $T_c$ reveals the opening of a spin gap in the superconducting state (Extended Data Fig.6). For comparison, spin excitations at $E=1.075\pm 0.025$ meV are clearly enhanced below $T_c$ at BT [Figs. 4(c,d)]. Figure 4(e) and (f) shows ${\bf Q}$ cuts along the $[0,K,0]$ direction at $E=0.25\pm 0.05$ and $E=1.05\pm 0.05$ meV, respectively. Although superconductivity in UTe$_2$ induces a spin gap and a resonance, it does not change the ${\bf Q}$-dependent lineshape, as seen in the resonance of CeCoIn$_5$ \cite{Song2020}. Figure 4(g) shows energy-dependent scattering at the Y1 point together with the nuclear incoherent scattering backgrounds taken at the background wavevector position ${\bf Q}_B$ [Fig. 1(c)]. We find clear evidence of a spin gap at BT below $E=0.25\pm 0.05$ meV and a resonance at $E_r=7.9k_BT_c$. Figure 1(e) and 1(f) shows the temperature dependence of $\chi^{\prime\prime}(E)$ at the Y1 and T1 positions, respectively, in absolute units, obtained by subtracting the incoherent scattering backgrounds, correcting for the Bose population factor, and normalizing the magnetic scattering to a vanadium standard. We note that the magnitude of the magnetic scattering in UTe$_2$ is similar to that of iron-based superconductors \cite{Dai}. The temperature dependence of the spin gap and resonance is obtained by systematically subtracting the high-temperature data (the average of the $T=1.8$ and 2 K data) from those at lower temperatures [Fig. 4(h)]. At $T=1.5$ K, the temperature-difference plot shows no visible feature. On cooling further below $T_c$, we find clear evidence for negative and positive scattering in the temperature-difference plots arising from the opening of a spin gap and the emergence of a resonance, similar to other unconventional spin-singlet superconductors \cite{mignod,Wilson2006,Dai,Sato2001,Bernhoeft2000,CStock2008,OStockert2011}. Figure 4(i) shows similar temperature difference plots between BT and 2 K obtained at Y1 and Y2 with $E_i=3.32$ meV, again revealing the resonance at these equivalent positions. The absence of the resonance mode at T1 and T2 is shown in the temperature difference plots of Fig. 4(j). Finally, Figure 4(k) summarizes the temperature dependence of the scattering at Y1 for energies of $E_{gap}=0.45\pm 0.25$ and $E_r=1.15\pm0.45$ meV. It is clear that the intensity gain of the resonance below $T_c$ occurs at the expense of opening a spin gap at energies below it.

\noindent
{\bf Discussion} \newline
To summarize the INS results in Figs. 2-4, the temperature dependence of $\chi^{\prime\prime}(E)$ at Y1 and T1 is plotted in Figs. 1(e) and 1(f), respectively. In previous work, the energy of the resonance $E_r$ for unconventional spin-singlet superconductors was found to be proportional to the universal value $E_r=5.8k_BT_c$ \cite{Wilson2006} or the superconducting gap $\Delta$ \cite{GYu2009}. The values of $E_r/k_BT_c$ of spin-singlet heavy-fermion superconductors are well below the dashed line representing $E_r/k_BT_c=5.8$, whereas $E_r/k_BT_c$ for UTe$_2$ is well above the dashed line [Fig. 1(g)]. Assuming that UTe$_2$ has a superconducting gap of $\Delta=0.25$ meV \cite{Jiao2020}, $E_r/2\Delta\approx 2$ for UTe$_2$ is well above the expected universal dashed line of $E_r/2\Delta\approx 0.6$ [Fig. 1(h)] \cite{GYu2009}. Because the resonance energy is believed to be a direct measure of the electron-pairing strength, arising from the spin-singlet to spin-triplet excitations for spin-singlet superconductors \cite{Eschrig}, its observation in UTe$_2$ suggests that the system might also be a spin-singlet superconductor, in contrast to previous work \cite{Ran2019,Aoki2019,Ran2019b,Knebel2019,Sundar2019,Jiao2020,Nakamine2021,Hayes2020}. By comparing magnetic scattering intensity at the Y1, Y2, and Y3 equivalent points in reciprocal space in Figs. 2(c-f), we conclude that spin fluctuations are highly anisotropic in spin space with a large magnitude along the $a$-axis direction, thus suggesting the presence of a large spin-orbit coupling. Three $5f$ electrons of uranium in UTe$_2$ can display a dual localized and itinerant character similar to other U-based compounds \cite{Zwicknagl2003,Amorese2020}, and so superconductivity can arise from some itinerant electrons, whereas magnetism comes about from other, more localized electrons. In this picture, the presence of the AF resonance in UTe$_2$ at an energy so different from other spin-singlet superconductors could simply be a consequence of the weak coupling between itinerant and localized electrons [Figs. 1(g,h)].

Alternatively, if we assume that UTe$_2$ is indeed a spin-triplet superconductor, our results reveal several important conclusions for the microscopic origin of spin-triplet superconductivity. First and foremost, the discovery of a resonance in UTe$_2$ with $E_r\approx 7.9k_BT_c$ suggests that AF spin fluctuations with large spin-orbit coupling can drive spin-triplet superconductivity, clearly different from the current understanding that FM spin fluctuations are responsible for its superconductivity \cite{Ishizuka2021}. Second, the observation of a superconductivity-induced spin gap at energies below $E_r$ suggests that the superconducting order parameter may have a spin-singlet component with a sign change (possibly in the $A_g$ state) \cite{Ishizuka2021}. Third, within a spin exciton picture of the resonance,  we expect $E_r({\bf Q})< \min(\left|\Delta({\bf k})\right|+\left|\Delta({\bf k}+{\bf Q})\right|)$ \cite{scalapinormp,Eschrig}. Because scanning tunnelling microscopy experiments reveal a superconducting gap of $\Delta=0.25$ meV \cite{Jiao2020}, $E_r/2\Delta\approx 2$ in UTe$_2$ is much larger than $E_r/2\Delta=0.64$ found in unconventional spin-singlet superconductors [Fig. 1(h)] \cite{GYu2009}. Fourth, our experimental observation of AF spin excitations extending up to about $E = 6$ meV in Fig. 3 suggests that the in-plane magnetic exchange coupling of UTe$_2$ has an energy scale about ten times the superconducting pairing energy of $2\Delta=0.5$ meV, similar to copper oxide \cite{scalapinormp} and iron-based superconductors \cite{Dai}. A crude estimation using Figs. 1(b,e) and 3(c) suggests that the saving of magnetic exchange energy in the superconducting state in UTe$_2$ is sufficient to account for the superconducting condensation energy determined from the heat capacity anomaly across $T_c$ (Extended Data Fig.2) \cite{MWang2013}. Finally, the discovery of a resonance and normal-state spin excitations in UTe$_2$, where charged quasiparticles can also be probed by angle-resolved photoemission spectroscopy \cite{wray2020}, should open new avenues of research towards understanding the connection between spin excitations and Fermi surface topology in UTe$_2$. 

UTe$_{2}$ is a multi-band/orbital system, with superconducting pairing channels classified by the $D_{2h}$ point group \cite{Hayes2020}. In such multi-band/orbital superconductors, the presence of additional orbital degrees of freedom expands the pool of symmetry-allowed spin-triplet and spin-singlet pairing candidates, which, in turn, implies that quasi-degenerate pairing channels are more probable than in the typical single-band cases. As outlined in Methods, AF spin correlations of UTe$_{2}$, in the probable $\Gamma_5$ $f$-ground manifold \cite{Amorese2020}, allow for not only the usual spin-singlet pairing channel but also spin-triplet pairing channels. The spin-triplet pairing channels arise because the product of two $\Gamma_5$'s contains not only the spin-singlet matrix $\Gamma_1$ but also three spin-triplet matrices $\Gamma_{2-4}$, which transform as three one-dimensional non-trivial representations. Therefore, spin-triplet pairing states are allowed by the AF correlations in the manifold of U $\Gamma_{5}$ doublets. Furthermore, in the presence of AF correlations and when the spin-orbit coupling induces strong Ising anisotropy, as is the case for UTe$_{2}$ \cite{Ran2019},  the spin-triplet channel can become energetically competitive.



\newpage
\begin{figure}[ht]
    \centering
    \includegraphics{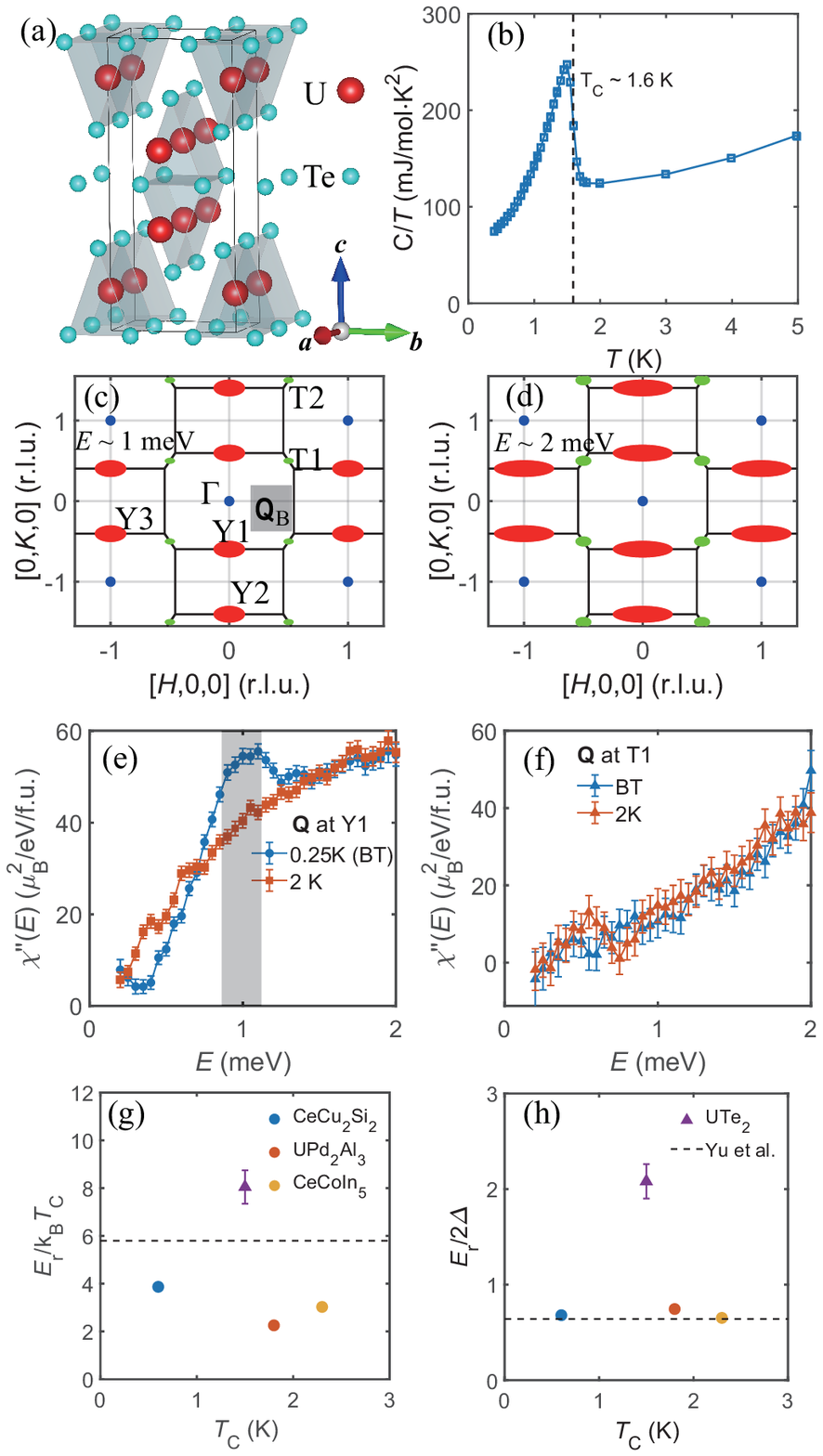}
    \caption{{\bf Crystal structure, heat capacity, and a summary of INS results.}}
    \label{fig:1}
\end{figure}
\begin{figure}[ht]
    \centering
    \includegraphics{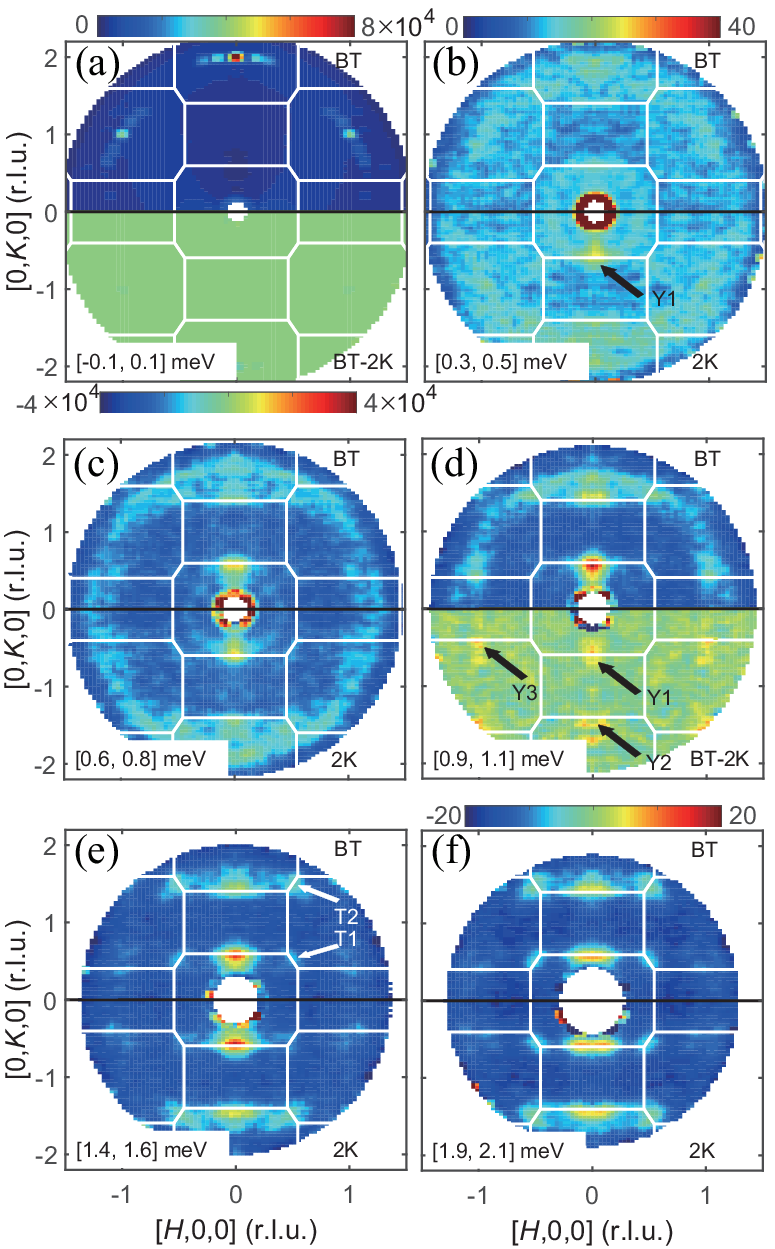}
    \caption{{\bf The wavevector, energy, and temperature dependence of the scattering function $S(\mathbf{Q},E)$ in the $[H,K,0] $ plane.}  }
    \label{fig:2}
\end{figure}
\begin{figure}[ht]
    \centering
    \includegraphics{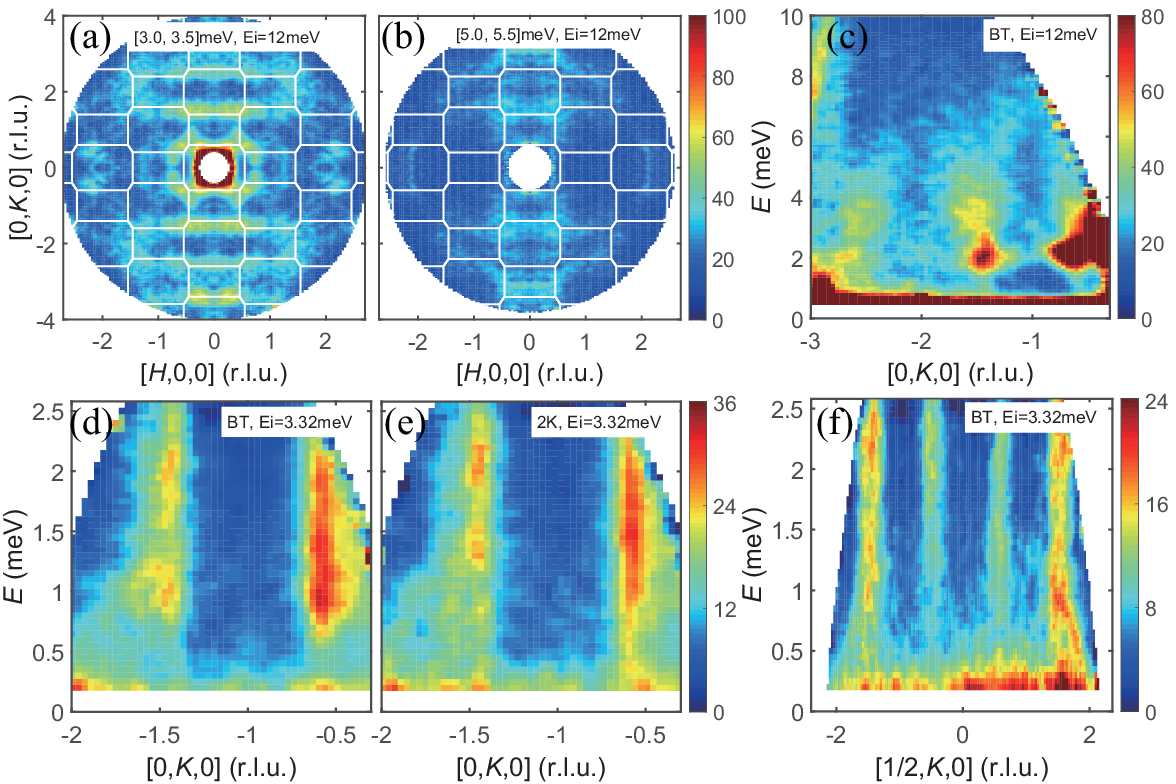}
    \caption{{\bf The wavevector and energy dependence of the scattering below and above $T_c$.}}
    \label{fig:3}
\end{figure}
\begin{figure}[ht]
    \centering
    \includegraphics{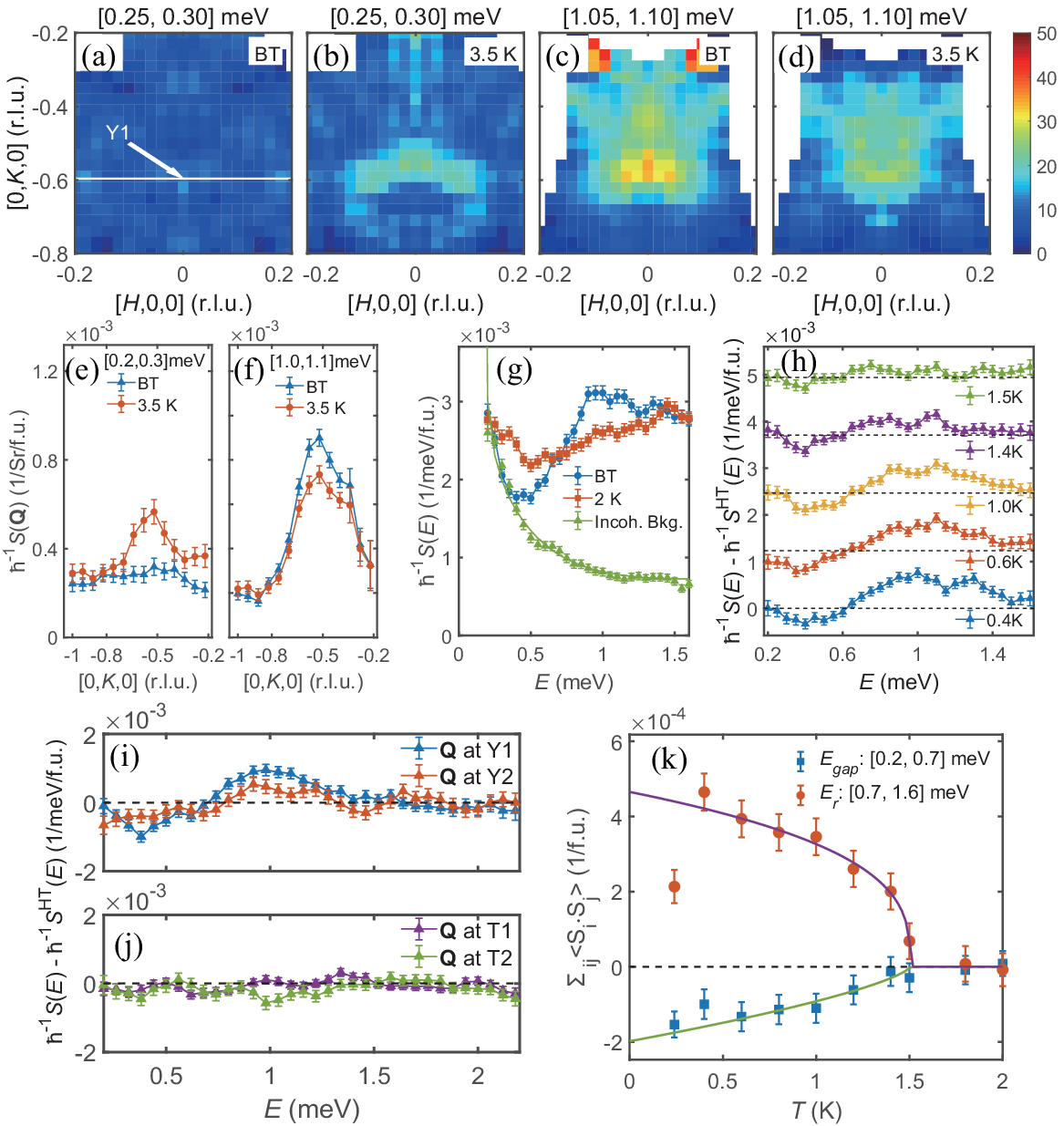}
    \caption{{\bf The wavevector, energy, and temperature dependence of the scattering at Brillouin zone boundary points.}}
    \label{fig:4}
\end{figure}

\begin{figure}[h]
\renewcommand{\thefigure}{1}
\caption{{\bf Crystal structure, heat capacity, and a summary of INS results.}
(a) The crystal structure of UTe$_2$. (b) The heat capacity data plotted as a function of temperature. 
A clear jump is observed at $T_c\approx 1.6$ K. (c,d) 
Schematic plots of the INS pattern in the $[H,K,0]$ plane at $E=1$ and 2 meV, respectively. 
Brillouin zones are marked with solid black lines, and the $\Gamma$ points are marked with blue dots. Spin excitations are observed 
at high symmetry points at Brillouin zone boundaries. The excitations at Y1 ($K=0.59$, $H=0$) and its symmetry equivalent positions 
(red ellipses, Y2 is at $K=1.41$) are coupled to superconductivity, as shown in the $\chi^{\prime\prime}(E)$ plot in (e), where the data taken at BT and 2 K are 
integrated in a box of $H$: $\pm0.12$, $K$: $\pm0.18$, $L$: $\pm0.5$ r.l.u. at the Y1 position with an $E$ step of 0.05 meV. 
A background taken at $\mathbf{Q}_B$ is subtracted from the integrated data to remove the incoherent scattering 
before making the Bose factor correction. $\mathbf{Q}_B$ has the same $|Q|$ with Y1 and is away from nuclear 
Bragg peaks and phonon modes. The background integration range is $H$: [0.22, 0.58], $K$: [-0.27, 0.27], $L$: [-0.75, 0.75] r.l.u., as marked 
in (c) with the shaded box. The $\chi^{\prime\prime}(E)$ at the T1 position is plotted in (f) as a comparison, which does not couple to superconductivity. (g,h) Comparison of the resonance energy of UTe$_2$ 
(this work) and its ratio to the superconducting energy gap \cite{Jiao2020} with other heavy-fermion superconductors: 
CeCu$_2$Si$_2$ \cite{OStockert2011}, UPd$_2$Al$_3$ \cite{Sato2001}, CeCoIn$_5$ \cite{CStock2008,Song2020} and the universal relationship summarized in Ref. \cite{GYu2009}. The vertical 
error bars in (e,f) represent statistical errors of 1 standard deviation. The vertical error bars in (g,h) represent uncertainty in $E_r/k_BT_c$ and $E_r/2\Delta$, respectively.
}
\end{figure}

\begin{figure}[h]
\renewcommand{\thefigure}{2}
\caption{ {\bf The wavevector, energy and temperature dependence of the scattering 
function $S(\mathbf{Q},E)$ in the $[H,K,0] $ plane.}  
Constant energy cuts of the symmetrized $S(\mathbf{Q},E)$ with $E_i = 3.32$ meV in  
(a) elastic channel, $E=0\pm 0.1$ meV; (b) 0.4 meV; (c) 0.7 meV; (d) 1.0 meV; (e) 1.5 meV; and (f) 2.0 meV. 
Unsymmetrized raw data are available in Extended Data Fig.4. The bin size is 0.035 r.l.u. along 
both the $H$ and $K$ directions. The integration range is $\pm$0.2 r.l.u. in $L$, and 
$\pm$0.1 meV in $E$. In each subplot, the upper panel shows data taken at BT. 
For (a) and (d), the lower panel shows the subtraction of data taken at BT and 2 K. 
For (b), (c), (e) and (f) the lower panel shows data taken at 2 K. High symmetry positions at 
the Brillouin zone boundaries (Y1, Y2, Y3, T1, T2) are marked with arrows in 
(b), (d) and (e). The unit of the color bars in Figs. 2, 3, 4(a-d)  
is $(\frac{1}{2}g\gamma_n r_0)^{-2}$mBarn/meV/Sr/f.u.,
where $g=2$ and $\frac{1}{2}\gamma_n r_0=2.695\times10^{-15}$ m. The colour bars above and below (a) are for the upper and lower panel of (a), respectively. The colour bar for (b), (c), upper panel (d), (e) and (f) is shown in (b). The colour bar in lower panel (d) is shown below (d).
 }
\end{figure}

\begin{figure}[h]
\renewcommand{\thefigure}{3}
\caption{ 
{\bf The wavevector and energy dependence of the 
scattering below and above $T_c$.}
(a,b) Constant energy cuts of the symmetrized scattering function 
$S(\mathbf{Q},E)$ with $E_i$ = 12 meV at $E=3.25$ meV and 5.25 meV at BT, respectively. 
Unsymmetrized raw data are available in Extended Data Fig.4-6. 
The bin size is 0.04 r.l.u. along both the $H$ and $K$ directions. 
The integration range is $\pm$0.2 r.l.u. in $L$, and $\pm$0.25 meV in $E$. (c) $E$-$\mathbf{Q}$ plot of $S(\mathbf{Q},E)$ 
at $H = 0$ with $E_i = 12$ meV at BT. The integration range is $\pm 0.2$ r.l.u. in $H$ and $\pm 0.3$ r.l.u. in $L$, the bin size along the $K$ 
direction is 0.04 r.l.u., and the $E$ step is 0.1 meV. The flat band near $E = 2$ meV is due to multiple scattering from the sample environment 
with $E_i = 12$ meV, and is absent in data with $Ei = 3.32$ meV. (d,e,f) $E$-$\mathbf{Q}$ plots of $S(\mathbf{Q},E)$ 
with $E_i = 3.32$ meV. (d) $H = 0$ at BT, (e) $H = 0$ at 2 K, and (f) $H = \frac12$ at BT. The integration range is 
$\pm 0.1$ r.l.u. in $H$ and $\pm 0.3$ r.l.u. in $L$, the bin size along $K$ is 0.035 r.l.u., and the $E$ step is 0.05 meV. The two strong intensity regions around 1.3 meV and 2 meV in (e) are statistical fluctuations (See cuts in Extended Data Fig.7), as they
do not appear in the data with $E_i=2.5$ meV in Extended Data Fig.6(f).  
 }
\end{figure}

\begin{figure}[h]
\renewcommand{\thefigure}{4}
\caption{ {\bf The wavevector, energy, and temperature dependence of the 
scattering at Brillouin zone boundary points}. (a-d) Constant energy cuts of the symmetrized scattering function $S(\mathbf{Q},E)$ with 
$E_i = 2.5$ meV at (a) $E=0.25$ to 0.3 meV and BT; (b) $E=0.25$ to 0.3 meV and 3.5 K; (c) $E=1.05$ to 1.1 meV and BT, and (d) $E=1.05$ to 1.10 meV and 3.5 K. The bin size is 0.02 r.l.u. in $H$ and 0.03 r.l.u. in $K$. The integration range is $\pm$0.3 r.l.u. in $L$. (e,f) One-dimensional (1D) cuts of $S(\mathbf{Q})$ with $E_i = 2.5$ meV across Y1 along the $K$ direction integrated in (e) $E=0.2$ to 0.3 meV and (f) $E=1.0$ to 1.1 meV at BT (blue triangles) and 3.5 K (red circles), respectively. The bin size is 0.06 r.l.u. in $K$. The integration range is $\pm$0.08 r.l.u. in $H$ and $\pm$0.3 r.l.u. in $L$. (g) 1D cuts of $S(\mathbf{Q})$ with $E_i = 3.32$ meV at Y1 along $E$ at BT (blue circles) and 2 K (red squares), respectively. Incoherent background scattering 
integrated at $\mathbf{Q}_B$ is plotted in green triangles. (h) 1D cuts of $S(\mathbf{Q})$ 
with high temperature data ($S^{HT}(\mathbf{Q})$) subtracted. The cuts are taken at Y1 along $E$ taken at 0.4 K (blue), 0.6 K (red), 1.0 K (yellow), 1.4 K (purple), and 1.5 K (green) with $E_i = 3.32$ meV. The high temperature data are averaged over 1.8 K and 2 K data, both above $T_c$. 
Different temperature data in (h) are artificially shifted, with the dashed black line representing the base line for each temperature. 
The integration ranges in all the 1D cuts in (g) and (h) are: 
$\pm$0.1 r.l.u. in $H$, $\pm$0.15 r.l.u. in $K$, and $\pm$0.3 r.l.u. in $L$. The bin size in $E$ is 0.05 meV. 
(i, j) Temperature difference $S(E)$ at different Brillouin zone points. The integration is done around each Y and T points in a box of $H$: $\pm0.1$, $K$: $\pm0.15$, $L$: $\pm0.5$ r.l.u. with an $E$ bin size of 0.06 meV. $E_i = 3.32$ meV. (k) $\chi^{\prime\prime}(E)$ versus temperature at $E_{gap}$ (blue squares) and $E_r$ (red circles). The integration is done around Y1 in a box of $H$: $\pm0.1$, $K$: $\pm0.15$, $L$: $\pm0.3$ r.l.u.. The solid line is a guide to the eye.  The y-axes in (e-j) are divided by the reduced Planck constant $\hbar=h/2\pi$ to have a dimension of $E^{-1}$. 
The vertical error bars in (e-k) represent statistical errors of 1 standard deviation. 
}
\end{figure}
\clearpage
\newpage
\pagebreak

\noindent
{\bf Methods} 
\noindent
{\bf Single crystal growth} Single crystals of UTe$_2$ were produced using an iodine vapor transport method similar to that 
described earlier \cite{Ran2019}. U (99.98\% purity) and Te (99.99\% purity) were combined in the ratio $2:3$ and sealed with iodine (3 mg/cm$^3$, 99.999\% purity) in an evacuated quartz tube with a length of 10 cm and an inner diameter of 1.4 cm. The tubes were placed in a single zone furnace with the hot end (furnace center) held at 1060$^\circ$C for four weeks. The natural temperature gradient of the furnace was adequate to promote vapor transport and to produce large single crystal specimens of the type shown in Extended Data Fig.1(a). After the heating cycle, samples were naturally cooled to room temperature, removed from the quartz tube, and rinsed in ethanol. Samples were subsequently stored under vacuum in sealed quartz ampoules.

\vspace*{0.25in}

\noindent
{\bf Heat capacity measurements} The temperature dependent heat capacity divided by temperature $C/T$ 
for two samples is shown in Extended Data Fig.2(a). Similar to earlier reports, there is a second order phase transition near $T_c\approx 1.6$ K, which marks the onset of superconductivity. All samples measured from these growth experiments show this feature but,  as previously reported, some show a single transition whereas others exhibit a double transition \cite{Hayes2020}. At higher temperatures, the data follow a Fermi liquid temperature dependence $C/T = \gamma+ \beta T^2$, 
where $ \gamma\approx 111$ mJ/mol-K$^2$, consistent with earlier reports \cite{Ran2019,Aoki2019}. 
We also find that the quantity $\Delta C/\gamma T_c \approx 1.54$, when 
it is determined using an equal entropy construction [Extended Data Fig.2(b)]. 
In order to estimate the superconducting condensation energy, 
we consider the expression $U(0) = 1/2N(0)\Delta^2(0)$ where $N(0)$ 
is the density of states 
at the Fermi energy and is determined from the expression 
$\gamma= \pi^2/3 k_B^2 N(0)$ and $2\Delta=3.52k_BT_c$ 
relates the BCS superconducting energy gap to the transition temperature. 
From this, we estimate 
$U(0) \approx 150$ mJ/mol, which is consistent 
with trends that are seen for other strongly correlated uranium based superconductors \cite{Kim2015}.  

\vspace*{0.25in}

\noindent
{\bf Neutron Scattering} INS measurements on UTe$_2$ were carried out using the Cold Neutron Chopper Spectrometer (CNCS)
 at Oak Ridge National Laboratory \cite{Ehlers2011}. 
The momentum transfer ${\bf Q}$ in three-dimensional reciprocal 
space is defined as ${\bf Q}$ = $H{\bf a}^* + K{\bf b}^* + L{\bf c}^*$, where $H, K$ and $L$ are Miller indices and ${\bf a}^* = \hat {\bf a}2\pi/a$, ${\bf b}^* = \hat {\bf b}2\pi/b$, ${\bf c}^* = \hat {\bf c}2\pi/c$ with $a$ = 4.16 \AA, $b$ = 6.12 \AA\ and $c$ = 13.95 \AA\ of the orthorhombic lattice \cite{Ran2019}. 
The crystals are naturally cleaved along the $ab$-plane and form small flakes about 0.5-1 mm thick and up to 1 cm long.  
We co-aligned 27 pieces (total mass 0.9-g) of single crystals on oxygen-free Cu-plates using an X-ray Laue machine to check
the orientation of each single crystal [Extended Data Fig.1(b,c) and 3].  The crystal assembly is aligned in the
$[H,K,0]$ scattering plane as shown in Figs. 1(c,d) and mounted on a $^3$He-insert installed in the standard cryostat. 
The lowest temperature can be reached in this setup is BT$=0.25$ K.
INS data were collected with incident neutron energies set to $E_i=12, 3.32,$ and 2.5 meV in the Horace mode as 
specified in the Figure 
captions \cite{RAEwings}. 
The sample co-alignment resulted in three assembled peaks at each Bragg positions with a 6-degree spread, as shown in Extended Data Fig.4(a,b). The strongest peak of the three contributes over 70\% of the total Bragg peak intensity. The $H, K$ coordinates used for data analysis was based on the position of the strongest peak. The symmetrized constant energy cuts shown in Figs. 2 (a-f), Figs. 3 (a,b), and Figs. 4(a-d) are results of two reflection operations against the horizontal and vertical axes according to the space group $Immm$, which do not change the positions of the strongest assembled peaks but create copies of the two smaller peaks on the opposite side. The symmetrized data were only used for constant energy cuts. All the one dimensional data shown in the main text and the SI are taken from the unsymmetrized raw data.  
Extended Data Figures 4, 5, and 6 show raw data obtained with $E_i=3.32$, 12, and 2.5 meV, respectively, at different 
temperatures. Extended Data Fig.7 shows cuts around FM Bragg peak and background positions 
at BT and 2 K, indicating no evidence
of FM spin fluctuations in UTe$_2$ within our measurement sensitivity.  We also checked possible existence
of quasielastic magnetic scattering, as seen in AF ordered UPd$_2$Al$_3$ \cite{Bernhoeft2000}, and find no
evidence in UTe$_2$ consistent with no static magnetic order in the system. 
 The high-flux instrument 
mode was used to maximize the neutron intensity with the Fermi chopper and double-disk 
chopper frequency at 60 Hz and 300 Hz, respectively.  The neutron scattering data are normalized to absolute units using a vanadium standard, which 
has an accuracy of approximately 30\%.

\vspace*{0.25in}

\noindent
{\bf Theory} In UTe$_{2}$, as is typical of $f$-electron materials, the U atomic states are split by strong spin-orbit-coupling  
and crystal-field effects into multiplets which transform according 
to the double-valued irreducible representations of the $D_{2h}$ point-group. 
We will construct microscopic Cooper pairing candidates of well-defined symmetry from products 
of momentum-dependent form-factors such as $p$-waves and matrices defined in the relevant multiplet space \cite{Nica_Si}.
The pairing matrices, which are obtained from the decomposition of the products of two multiplets, 
also transform as irreducible representations of the point group. 
This classification naturally restricts the number of symmetry-allowed pairing states 
by incorporating the spin-orbit-coupling and crystal-field  splitting for the U levels. 
Furthermore, by taking into account the relevant atomic structure of the paired electrons, 
pairing candidates constructed from our microscopic procedure go beyond the more common Landau-Ginzburg analysis, which relies only on a symmetry classification without reference to 
the pairing matrix structure. Our approach also
 provides a natural link to the topology of the superconducting state.

In this approach, the matrix structure in orbital/spin space \cite{Nica_Yu_Si}, or similarly in multiplet space with strong spin-orbit-coupling, provides the key to advancing new pairing states. To set the stage, we recall the  approach in the previously studied case \cite{Nica_Si}
of the prototypical heavy-fermion unconventional superconductor CeCu$_2$Si$_2$. In that compound,
various probes \cite{Nica_Si} point toward a ground-state $\Gamma_{7}$ Kramers doublet of the $D_{4h}$ point group which emerges from the Ce $f$-electron via spin-orbit-coupling and crystal-field splitting. Ref.~\cite{Nica_Si} showed that the matrix corresponding to spin-singlet pairing between two $\Gamma_{7}$ $f$ electrons transforms as the identity ($\Gamma_{1}$) irreducible representation of $D_{4h}$, which is featureless in the sense that it can be classified entirely via the symmetry of its form factor. However, the same procedure also predicted that, when paired instead with $\Gamma_{6}$ conduction electrons originating  from the Cu $d-$electron states, the $\Gamma_{7}$ $f$-electron multiplets give rise to a spin-singlet matrix that transforms as a $\Gamma_{3}$ irreducible representation; it changes signs under $C_{4z}$ rotations and thus transforms non-trivially. In CeCu$_{2}$Si$_{2}$, this pairing matrix, together with a featureless $s$-wave form factor, is equivalent to an unconventional $d+d$ pairing state consisting of intra- and inter-band $d$-wave components~\cite{Nica_Yu_Si}, reflecting the sign-changing nature of the irreducible representation. The $d+d$ pairing leads to a fully gapped Bogoliubov-de Gennes spectrum at lower temperatures. This matrix pairing state proved successful in accounting for the spin resonance observed 
 in CeCu$_{2}$Si$_{2}$ in inelastic neutron-scattering, as well as in fitting the experimental data 
 on London penetration-depth and specific-heat measurements that encode a hard gap in the low-energy BdG spectrum~\cite{Pang}. A small but nonzero
 admixture of $\Gamma_{6}$ $f$ electrons in the ground-state, as indicated by soft x-ray absorption spectroscopy~\cite{Amorese_PRB_2020}, provides evidence for the degrees of freedom that underlie the proposed $d+d$ pairing.

Although less is known about the $f$-electron levels of
UTe$_{2}$ at this stage, we can still construct and classify symmetry-constrained pairing channels for this compound using the same microscopic framework. A number of available \textit{ab initio} studies \cite{Duan2020,Shick2019,Shick_Fujimori_Pickett_PRB_2021} point toward a predominant U $j=5/2, m_{j}= \pm 1/2$ doublet at low energies. These results are consistent with data from core-level photoelectron spectroscopy~\cite{Amorese2020}.
They are also compatible with the spin size extracted in this work: By assuming that the spin excitation spectral weight determined from the $E_i=3.32$ meV data goes up linearly as a function of $E$ up to $6$ meV (the band top, which is determined by high $E_i$ data),
we estimate the momentum- and energy-integrated spin spectral weight to be $\approx\ 1.5\ \mu_B^2$/U that,
for $g$ close to $2$, is compatible with a spin size $1/2$. The double-valued irreducible representations of $D_{2h}$ only allow for $\Gamma_{5}$ Kramers doublets~\cite{Koster}. It is then natural to identify a  
$\Gamma_{5}$ doublet with the U $j=5/2, m_{j}= \pm 1/2$ states.

We can then proceed along the lines set out in Ref.~\onlinecite{Nica_Si}, and determine 
the possible pairing matrices via a decomposition of the product of two $\Gamma_{5}$ doublets. 
The products decompose as follows~\cite{Koster}: $\Gamma_{5} \times \Gamma_{5} =  \Gamma_{1} + \Gamma_{2} + \Gamma_{3} + \Gamma_{4}$. (Note that the parity of the $\Gamma_{5}$'s is not specified, but it does not affect the decomposition.)
In the above decomposition, the first 
term corresponds to a spin-singlet matrix, which transforms according to the identity $\Gamma_{1}$ representation. This component captures the standard result, {\it viz.} AF correlations  promote spin-singlet pairing.

What our procedure also reveals is a striking new result: The decomposition also includes three spin-triplet matrices. The latter transform according to three one-dimensional, non-trivial $\Gamma_{2-4}$ representations. We achieve our key results:
AF correlations within the ground-state manifold of U $\Gamma_{5}$ can also lead to
spin-triplet superconducting pairing. We re-iterate that, in arriving at this conclusion, it is crucial to account for the matrix structure of the pairing state.

We next turn to the energetics of the pairing states. A systematic study requires the knowledge of the both the tight-binding parametrization of the noninteracting bands and the effective interaction parameters among the $\Gamma_5$ multiplets. When such parameters are known, we can determine and compare the ground-state energies of the different pairing channels,
in the same way as used to show that the band-mixing 
$d+d$ (matrix spin-singlet) pairing state is energetically competitive~\cite{Nica_Si,Nica_Yu_Si}.
Given that the model parameters are not yet available, we resort to more general means to assess the stability of the spin-triplet pairing. A key feature is that the spin-orbit-coupling of UTe$_2$ is such that the magnetic response is strongly Ising anisotropic \cite{Ran2019}. For antiferromagnetically correlated systems that are highly Ising anisotropic, the spin-triplet channel can be energetically competitive, as captured in the microscopic calculations of pairing correlations in well-defined Kondo systems~\cite{Pixley_2015} and recently discussed in the context of superconductivity observed near a magnetic-field-induced heavy-fermion quantum critical point~\cite{Nguyen2021}.

\noindent
{\bf Data availability} The data that support the plots within this paper and other findings of this study are available from the corresponding authors upon reasonable request.

\vspace*{0.25in}

\noindent
{\bf Acknowledgements} P.D. thanks Doug Natelson, W. P. Halperin, Nick Butch, and Johnpierre Paglione for helpful discussions. 
E. M. N. and Q. Si acknowledge useful discussions with H. Hu, S. Paschen, and J.-X. Zhu.
The INS work at Rice is supported by the U.S. DOE, BES 
under grant no. DE-SC0012311 (P.D.). Part of the material characterization efforts at Rice is supported by the Robert A. Welch
Foundation Grant Nos. C-1839 (P.D.). 
Work performed by R.E.B. at the National High Magnetic Field Laboratory  was supported  by  National Science Foundation Cooperative Agreement No.
DMR-1644779 and the State of Florida.  Synthesis of crystalline materials and measurements by R.E.B. were supported by the Center for Actinide Science and  Technology  (CAST),  an  Energy  Frontier  Research Center (EFRC) funded by the U.S. DOE, BES, under grant no.  DE-SC0016568.
Research at UC San Diego was supported by the
U.S. DOE, BES under grant no. DEFG02-04-ER46105 (single crystal growth) and U.S. NSF under Grant No. DMR-1810310 (characterization of
physical properties). The theory work at Rice has primarily been supported by the U.S. DOE, BES under Award No. DE-SC0018197, with travel support provided 
by the Robert A. Welch Foundation Grant No. C-1411. Q.S. acknowledges the hospitality of the Aspen Center for Physics, which is supported by the NSF grant No. PHY-1607611. E.M.N. was supported by ASU startup grant. A portion of this research
used resources at the Spallation Neutron Source, a DOE Office of Science User
Facility operated by ORNL.

\vspace*{0.25in}

\noindent
{\bf Author contributions} P.D. and M.B.M. conceived the project. R.E.B. grew the single crystals and made 
specific heat measurements on the crystals. The single crystals of UTe$_2$ were aligned using Laue x-ray diffraction by C.D., Y.D., C.M. and A.J.B. and characterized by means of powder x-ray diffraction by C.M., A.J.B. and Y.D. at UCSD. The INS experiments were carried out by A.P. in remote discussion with C.D. and P.D..  
The data analysis was carried out by C.D. and P.D..  E.M.N. and Q.S. contributed to the theoretical
idea that AF spin fluctuations may facilitate spin-triplet superconductivity. The paper
was written by P.D., C.D., R.E.B., E.M.N, and Q.Si. all coauthors made comments on the paper.

\vspace*{0.25in}

\noindent
{\bf Competing interests} The authors declare no competing interests.

\vspace*{0.25in}

\noindent
{\bf Additional information} Correspondence and requests for materials should be addressed to P.D.

\noindent
\newpage
{\bf Extended Data}
\renewcommand{\figurename}{Extended data Fig.}
\renewcommand{\thefigure}{1}
\begin{figure}[ht]
	\centering
	\includegraphics[scale = 1.8]{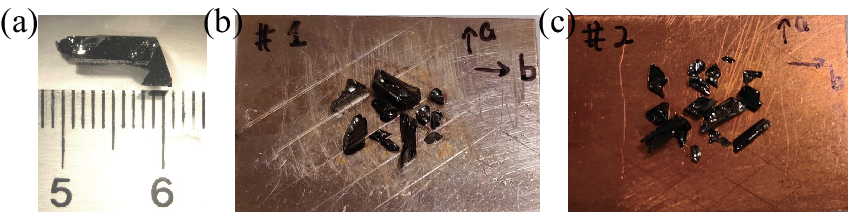}
	\caption{{\bf Pictures of the UTe$_2$ single crystals used in the INS experiment.} (a) A typical piece of UTe$_2$ single crystal of 10 mm by 3 mm by 3 mm in size. The direction of the longest edge is the intersection of $[1,1,0]$ plane and $[0,0,1]$ plane. 
	(b,c) 27 pieces of UTe$_2$ single crystals co-aligned on two oxygen-free Cu sample plates. The total mass is 0.9 gram.}
	\label{Fig:S1}
\end{figure}

\renewcommand{\thefigure}{2}
\begin{figure}[ht]
	\centering
	\includegraphics[scale = 0.5]{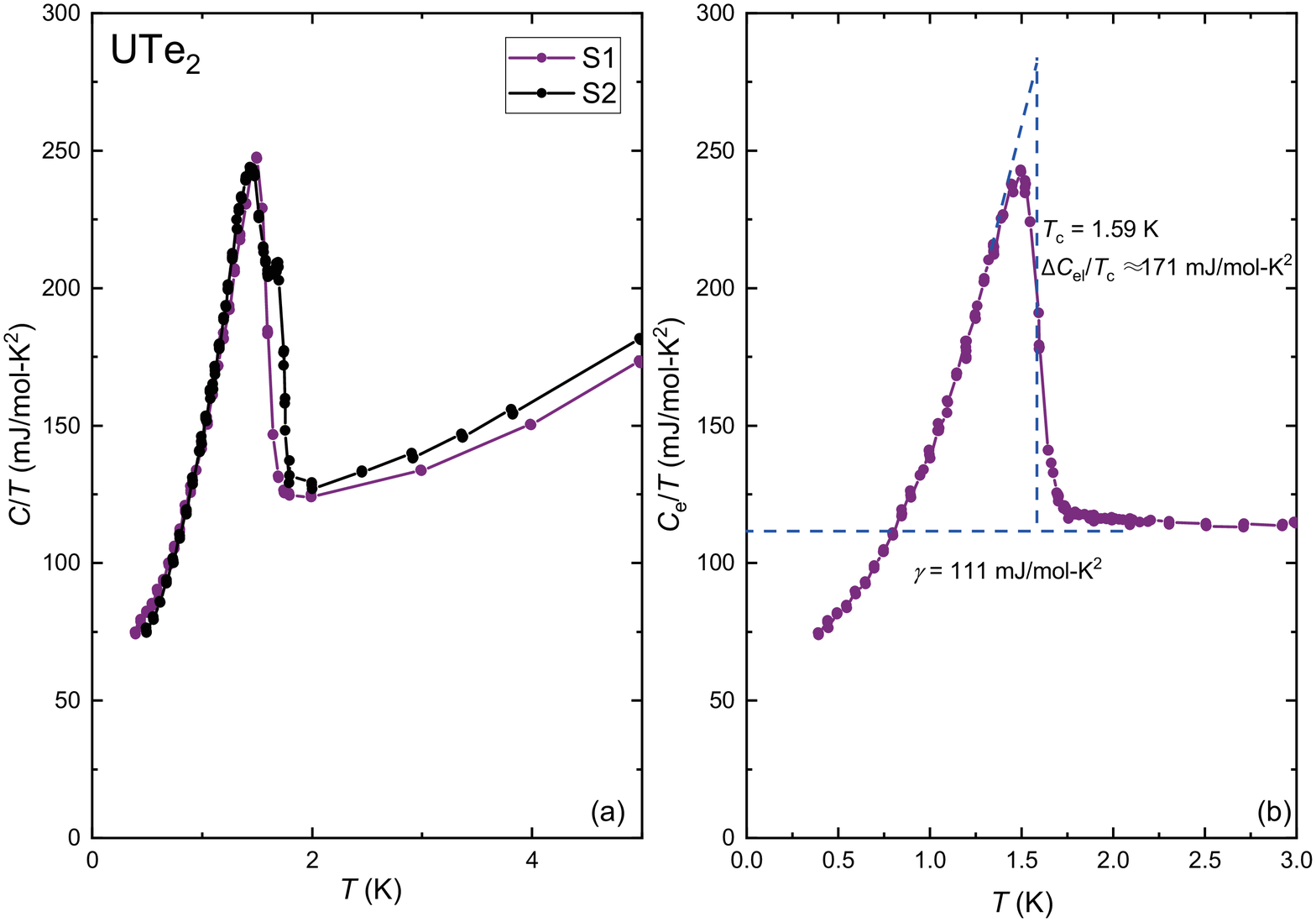}
	\caption{{\bf Summary of temperature dependent heat capacity $C/T$ for single crystal 
	specimens of UTe$_2$}. (a) Comparison of $C/T$ vs $T$ for two representative crystals 
	of UTe$_2$. One crystal shows a single superconducting phase transition 
	whereas the other shows two features. Several other crystals were measured, 
	which all show similar behavior. (b) The electronic component of the heat capacity $C_e/T$, 
	which was obtained by subtracting the low temperature phonon heat capacity $\beta T^2$, 
	which was obtained by fitting the data for $T > T_c$ 
	using the expression $C/T = \gamma + \beta T^2$. The normal state electronic 
	coefficient of the heat capacity $\gamma$ is indicated by the 
	horizontal dotted blue line. An equal entropy construction is also indicated by dotted blue lines 
	to determine $T_c$ and the ideal size of the heat capacity jump $\Delta C/T_c$.}
	\label{Fig:S2}
\end{figure}

\renewcommand{\thefigure}{3}
\begin{figure}[ht]
	\centering
	\includegraphics[scale = 1.]{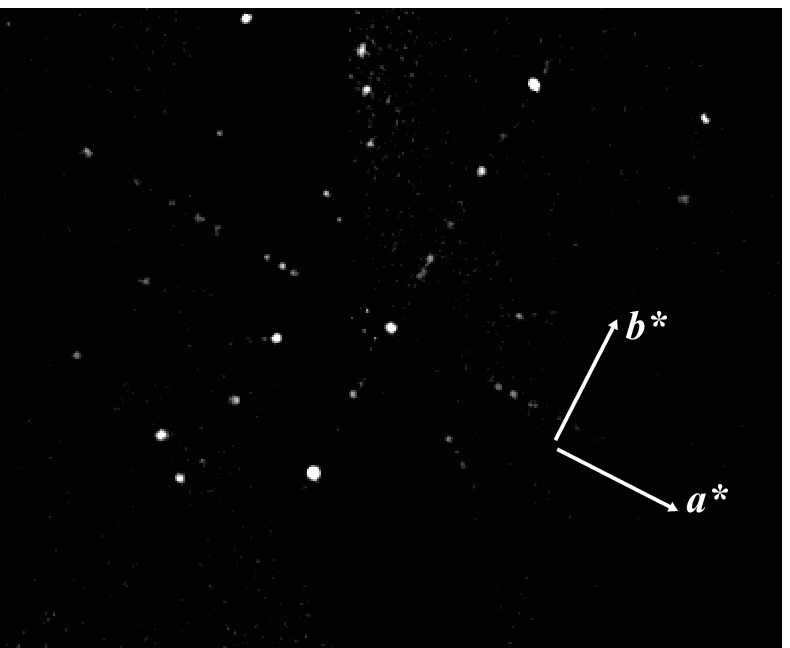}
	\caption{{\bf X-ray Laue pattern of the $[0,0,1]$ plane of UTe$_2$ for one of the samples used in the experiment.}}
	\label{Fig:S3}
\end{figure}

\renewcommand{\thefigure}{4}
\begin{figure}[ht]
	\centering
	\includegraphics[scale = 1.]{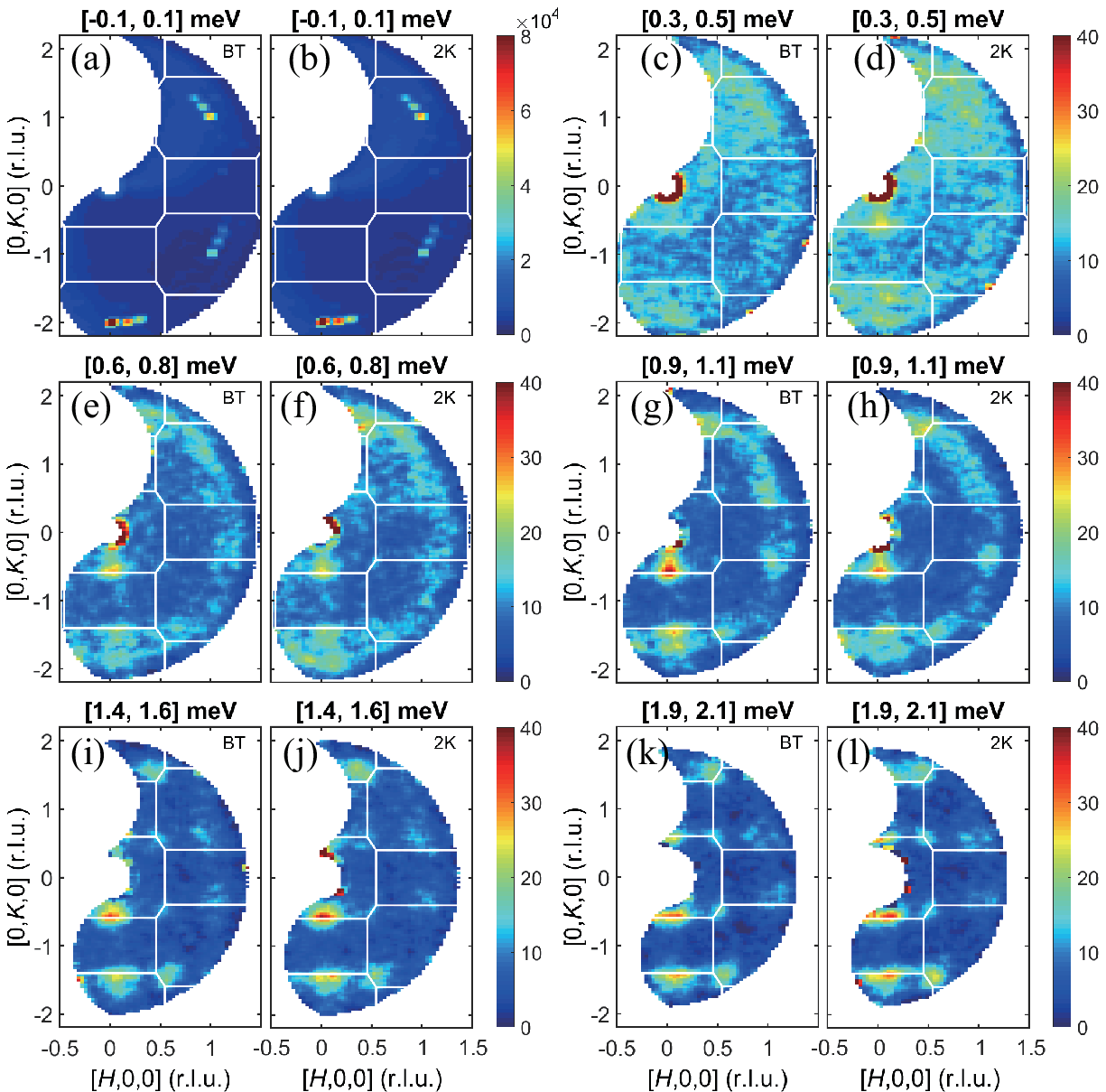}
	\caption{\textbf{Unsymmetrized raw data in the $[H,K,0]$ plane with $E_i$ = 3.32 meV.} Constant energy cuts of the unsymmetrized $S(\mathbf{Q},E)$ with $E_i$ = 3.32 meV at (a) $0.0\pm0.1$ meV and BT, (b) $0.0\pm0.1$ meV and 2 K, (c) $0.4\pm0.1$ meV and BT, (d) $0.4\pm0.1$ meV and 2 K, (e) $0.7\pm0.1$ meV and BT, (f) $0.7\pm0.1$ meV and 2 K, (g) $1.0\pm0.1$ meV and BT, (h) $1.0\pm0.1$ meV and 2 K, (i) $1.5\pm0.1$ meV and BT, (j) $1.5\pm0.1$ meV and 2 K, (k) $2.0\pm0.1$ meV and BT, (l) $2.0\pm0.1$ meV and 2 K. The bin size is 0.035 r.l.u. along both $H$ and $K$. The integration range is $\pm$0.2 r.l.u. in $L$, and $\pm$0.1 meV in $E$. The unit of the color bars in Supplementary Figs. 4, 5, 6 is the same as that of Fig. 2 of the main text.}
	\label{Fig:S4}
\end{figure} 

\renewcommand{\thefigure}{5}
\begin{figure}[ht]
	\centering
	\includegraphics[scale = 1.]{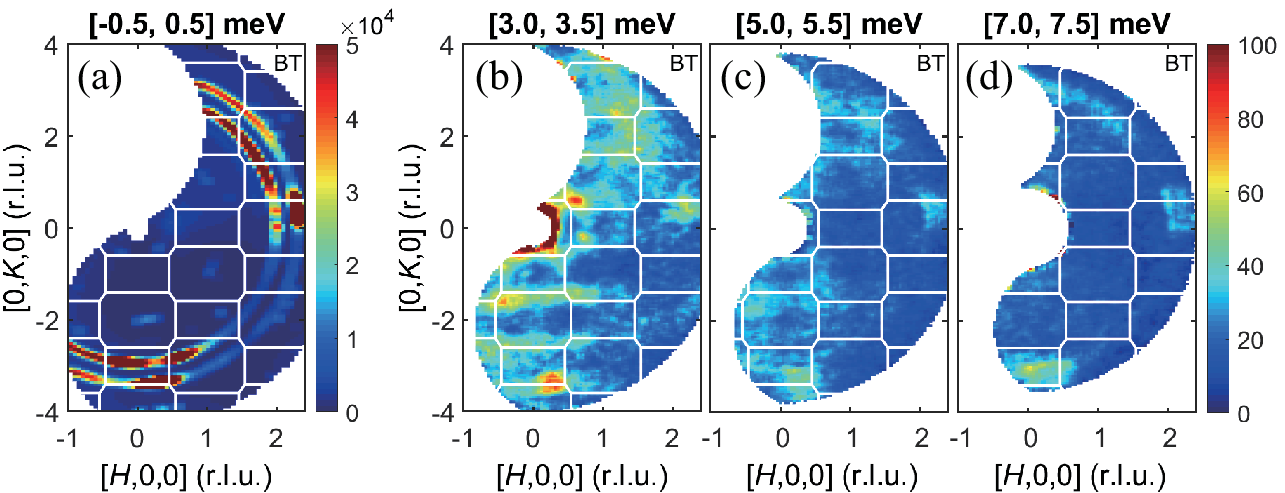}
	\caption{\textbf{Unsymmetrized raw data in the $[H,K,0]$ plane with $E_i$ = 12 meV.} Constant energy cuts of the unsymmetrized $S(\mathbf{Q},E)$ with $E_i$ = 12 meV and BT at (a) $0.0\pm0.5$ meV, 
	(b) $3.25\pm0.25$ meV, (c) $5.25\pm0.25$ meV, (d) $7.25\pm0.25$ meV. The bin size is 0.04 r.l.u. along both $H$ and $K$. The integration range is $\pm$0.2 r.l.u. in $L$, and $\pm$0.25 meV in $E$.  The rings of scattering in panel (a)
	are from the nuclear $(1,1,1)$ and $(2,0,0)$ Bragg peaks of the Cu sample holder.}
	\label{Fig:S5}
\end{figure} 

\renewcommand{\thefigure}{6}
\begin{figure}[ht]
	\centering
	\includegraphics[scale = 1.]{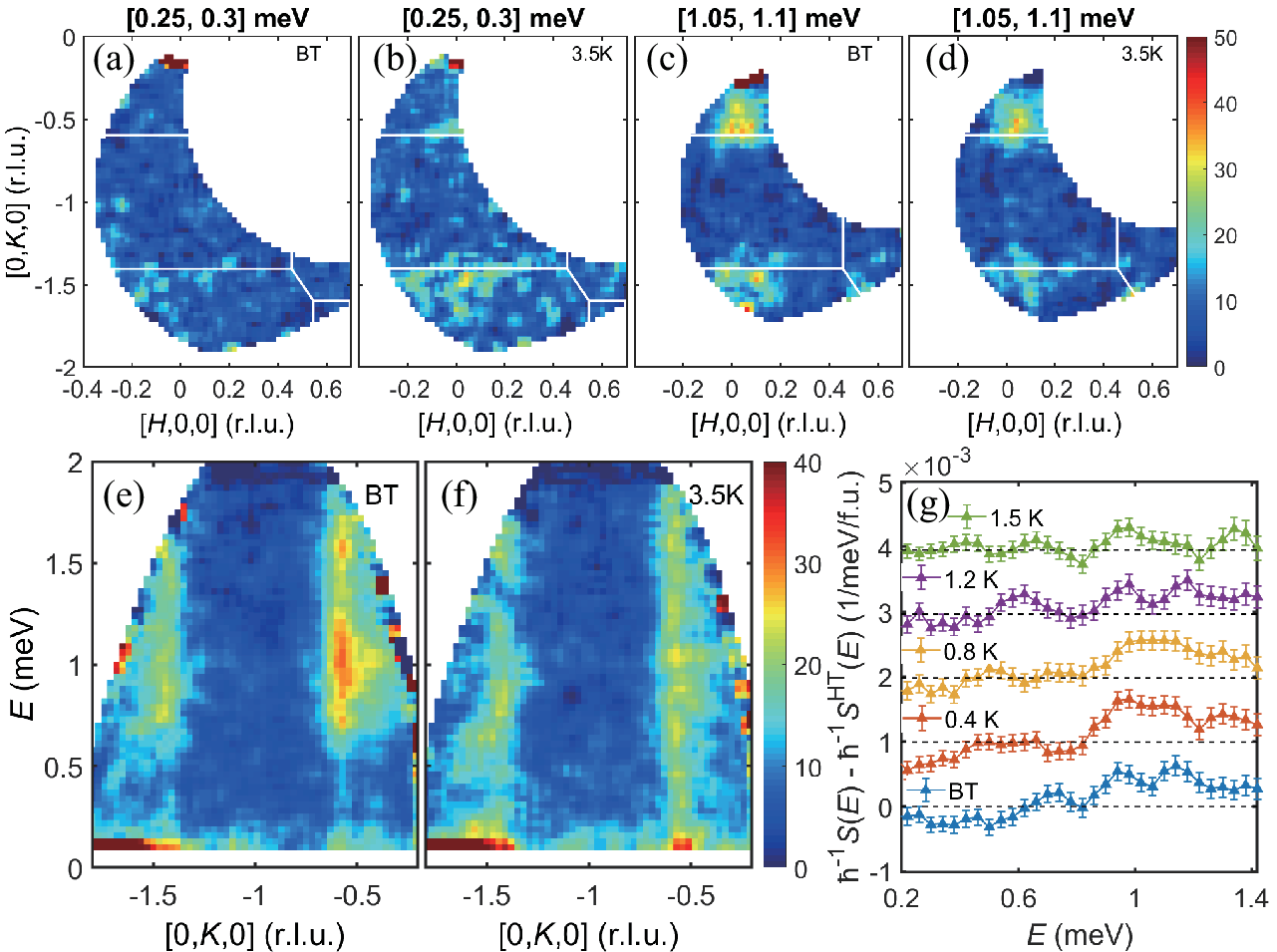}
	\caption{\textbf{Unsymmetrized raw data, $E$-$\mathbf{Q}$ plots and one-dimensional energy cuts with $E_i$ = 2.5 meV.} (a-d) Constant energy cuts of the unsymmetrized $S(\mathbf{Q},E)$ with $E_i$ = 2.5 meV at (a) 0.25 to 0.3 meV and BT, (b) 0.25 to 0.3 meV and 3.5 K, (c) 1.05 to 1.1 meV and BT, (d) 1.05 to 1.1 meV and 3.5 K. The bin size is 0.02 r.l.u. along $H$ and 0.03 r.l.u. along $K$. The integration range is $\pm$0.3 r.l.u. in $L$. (e,f) $E$-$\mathbf{Q}$ plots of the scattering function $S(\mathbf{Q},E)$ with $E_i$ = 2.5 meV at BT (e) and 3.5 K (f), respectively. The integration range is $\pm 0.08$ r.l.u. in $H$ and $\pm 0.3$ r.l.u. in $L$, the bin size along $K$ is 0.03 r.l.u., and the $E$ step is 0.03 meV. (g) one-dimensional cuts of the scattering function $S(\mathbf{Q})$ with high temperature data ($S^{HT}(\mathbf{Q})$) subtracted. The cuts are taken at Y1 along $E$ taken at BT (blue), 0.4 K (red), 0.8 K (yellow), 1.2 K (purple), and 1.5 K (green) with $E_i$ = 2.5 meV. The high temperature data is taken at 3.5 K. At low energy the excitation at Y1 is not fully covered with this $E_i$, which causes the gap feature between 0.2 to 0.7 meV hard to observe in the subtracted one-dimensional data. Different temperature data in (g) are artificially shifted, with the dashed black line representing the base line for each temperature. The integration ranges in (g) are: $\pm$0.08 r.l.u. in $H$, $\pm$0.15 r.l.u. in $K$, and $\pm$0.3 r.l.u. in $L$. The bin size in $E$ is 0.04 meV.}
	\label{Fig:S6}
\end{figure}

\renewcommand{\thefigure}{7}
\begin{figure}[ht]
	\centering
	\includegraphics[scale = 1.]{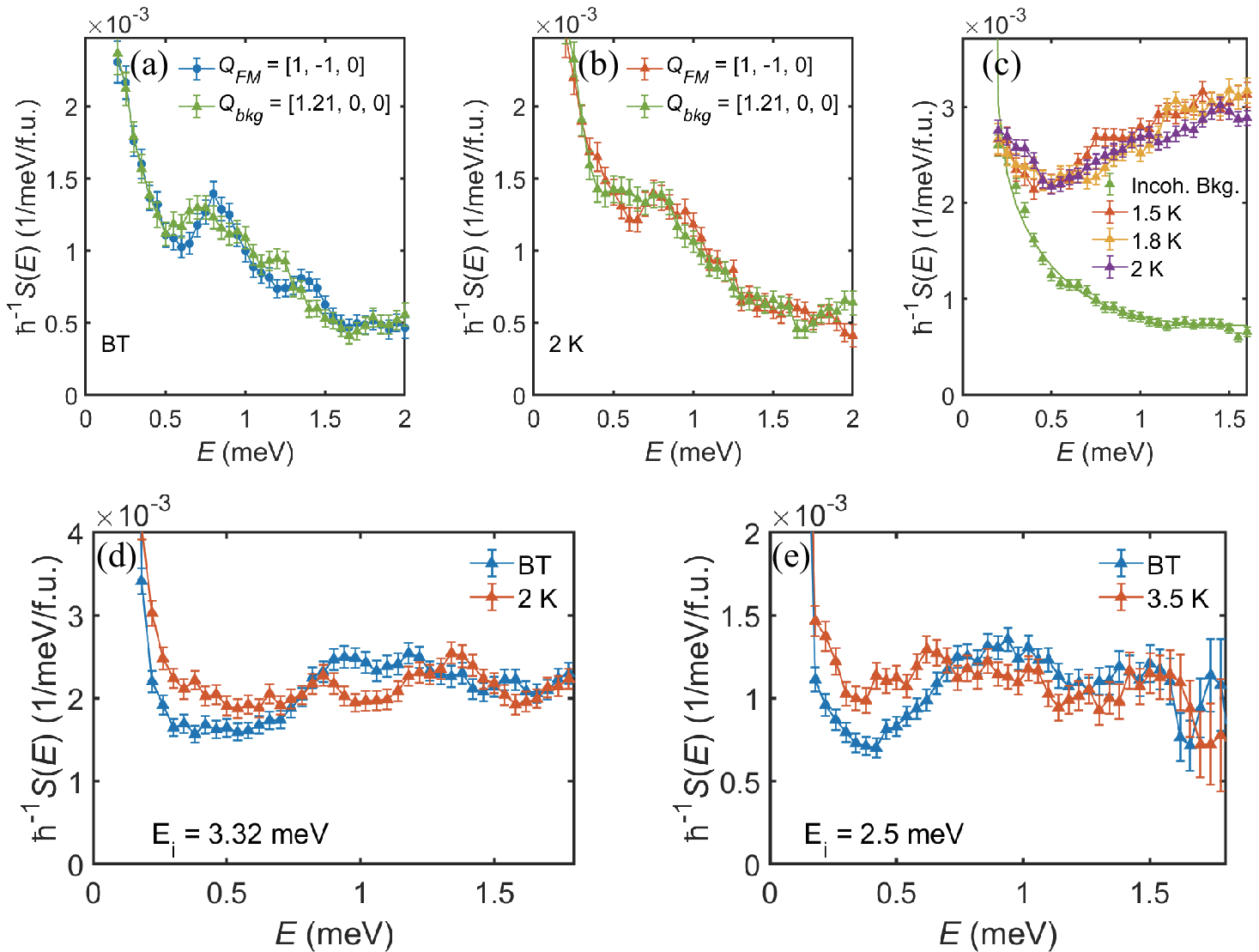}
	\caption{\textbf{Temperature dependence of the excitations at different $\mathbf{Q}$ positions} (a, b) One-dimensional cuts of $S(\mathbf{E})$ with $E_i = 3.32$ meV at Bragg peak $(1, -1, 0)$ along $E$ at BT and 2 K, respectively. Incoherent background scattering integrated at $\mathbf{Q}_{bkg}$ is plotted in green triangles. There are no FM spin fluctuation signals observed above the background. The broad peak around $E=0.7$ meV is powder
	ring of scattering not associated with UTe$_2$ [see Extended Data Fig.4(e,g)]. (c) One-dimensional cuts of $S(\mathbf{E})$ with $E_i = 3.32$ meV at Y1 along $E$ at 1.5, 1.8, and 2 K. There is no significant change in the quasielastic energy range for temperature close to and above $T_c$. (d,e) One-dimensional cuts of $S(\mathbf{E})$ with $E_i = 3.32$ meV (d) and $2.5$ meV (e), respectively. The subtle increase of $S(\mathbf{E})$ above $T_C$ near 1.4 meV with $E_i = 3.32$ meV is just above one standard deviation, and is not observed with $E_i = 2.5 meV$. The integration ranges of the one-dimentional data in (d,e) are: $\pm$0.1 r.l.u. in $H$, $\pm$0.15 r.l.u. in $K$, and $\pm$0.3 r.l.u. in $L$. The bin size in $E$ is 0.04 meV. }
	\label{Fig:S7}
\end{figure}

\end{document}